\documentclass[11pt]{article}
\usepackage[colorlinks=true, citecolor=blue, urlcolor=blue, linkcolor=blue, breaklinks=true, pdfpagelabels=false]{hyperref}
\usepackage{cite}

\usepackage{hyperref}

\usepackage{amsmath,amsfonts,amssymb}
\usepackage{graphicx}
\usepackage{epsfig,epsf}
\usepackage{bm}

\def\be {\begin{equation}}
\def\ee {\end{equation}}
\def\bea {\begin{eqnarray}}
\def\eea {\end{eqnarray}}

\def\barr{\begin{array}}
\def\earr{\end{array}}

\def\opcit(#1){ {\em op. cit.}, #1}

\def\issue(#1,#2,#3){#1, #2 (#3)} 

\def\equationautorefname~#1\null{Eq.\,(#1)\null}
\def\pageautorefname\nobreakspace{p.}

\makeatletter\renewcommand{\p@subsection}{\thesection.}\makeatother%

\begin{document}

\renewcommand*{\thefootnote}{\fnsymbol{footnote}}


\begin{center}
{\Large\bf{Charged Higgs Decay to $W^{\pm}$ and Heavy Neutral Higgs Decaying into $\tau^+\tau^-$ in Georgi-Machacek Model at LHC}}


\vspace{5mm}

{\bf Swagata Ghosh}$^{}$\footnote{swgtghsh54@gmail.com}

\vspace{3mm}
{\em{Department of Physics, Indian Institute of Technology Kharagpur, Kharagpur 721302, India.
}}

\end{center}

\begin{abstract}

The CMS collaboration at the Large Hadron Collider (LHC) searched for a charged Higgs boson, in the mass range of 300 to 700 GeV, decaying into a $W^{\pm}$ boson and a heavy neutral Higgs boson of mass $200$ GeV, which successively decays into a pair of tau leptons, in proton-proton collisions at $\sqrt{s} = 13$ TeV. In this letter, focusing on the Georgi-Machacek (GM) model, I discuss the parameter space, allowed by the theoretical and experimental constraints, for which the limits on this process obtained by the CMS can be accommodated. The study in this letter also shows that, for the choice of the parameters, the decay of the charged Higgs boson $H_3^{\pm}$ to $W^{\pm}$ and a heavy neutral Higgs boson H is preferred over the decay to any gauge boson and any other neutral or charged Higgs bosons. 
I also present the values of production cross-section times branching ratio for the decay of H to a pair of b-quarks at $\sqrt{s}=14$ TeV.

\end{abstract}



\setcounter{footnote}{0}
\renewcommand*{\thefootnote}{\arabic{footnote}}

\section{Introduction}
\label{intro}

The discovery of the Higgs boson at the LHC \cite{ATLAS:2012yve,CMS:2012qbp} brings the Standard Model (SM) into triumphant, though it cannot explain all of the natural observations and theoretical problems.
To resolve these deficiencies, one of the common practices is to add higher multiplets, like additional singlet, doublet, or triplet to the scalar sector of the SM.
In this letter, I am interested in a model where two scalar triplets, one real and one complex, are added to the SM, which is popularly known as the Georgi-
Machacek (GM) model \cite{Georgi:1985nv, Logan:2015xpa, Kundu:2021pcg}.
The custodial symmetry is preserved in this model, $i.e.$ $\rho=1$ \cite{Chiang:2012cn}.
The constraints on the parameter space of the model are already well studied in the literature \cite{Hartling:2014aga, Chiang:2014bia, Du:2022brr, Chiang:2013rua, Chiang:2018cgb, Ismail:2020zoz, Chen:2022zsh, Bairi:2022adc, Ghosh:2022bvz}, and also a calculator is available \cite{Hartling:2014xma} to calculate all the constraints.
Various phenomenological works \cite{Chiang:2015rva,Chiang:2014hia,Chiang:2015kka,Mondal:2022xdy,Chang:2017niy,Degrande:2017naf,Zhou:2018zli,Logan:2017jpr,Das:2018vkv,Ghosh:2019qie,Banerjee:2019gmr,Adhikary:2020cli,deLima:2022yvn} on this model enrich the knowledge of the reader.
The articles \cite{Ghosh:2022wbe,Chiang:2015amq,Cen:2018okf} explore the decay of the CP-odd singly charged Higgs boson, but the channel probed in this letter, first described in the paper of CMS experiment \cite{CMS:2022tkb}, is still untouched in the literature for the GM model.\\ \\
The mass $m_{H^{\pm}}$ of any singly charged Higgs boson $H^{\pm}$ can be split into three regions, heavy $(m_{H^{\pm}}>m_t-m_b)$, intermediate $(m_{H^{\pm}}\approx m_t)$, and light $(m_{H^{\pm}}<m_t-m_b)$, where $m_t$ and $m_b$ are top quark and bottom quark masses, respectively.
This letter addresses the production of heavy-charged Higgs, and as it is associated with top and bottom quarks at LHC, I also consider that production channel $i.e.$ $pp\rightarrow t b H^{\pm}$.\\ \\
The Georgi-Machacek model is formulated by the extension of the scalar sector of the SM by one real and one complex triplet with equal vacuum expectation value (vev) which results in the preservation of the custodial symmetry.
Besides the two CP-even neutral Higgs bosons $h$ and $H$, one CP-odd fermiophilic neutral Higgs $H_3$, and one CP-even but fermiophobic neutral Higgs $H_5$, GM model consists of four singly charged Higgs $H_3^{\pm},\,H_5^{\pm}$, and two doubly charged Higgs $H_5^{\pm\pm}$. $H_3,\,H_3^{\pm}$ are the members of the custodial triplet and $H_5,\,H_5^{\pm},\,H_5^{\pm\pm}$ are the members of the custodial quintuplet sharing the common mass $m_3$ and $m_5$, respectively.
The production cross-section of $H_5^{\pm}$ is much lower than that of $H_3^{\pm}$ and hence, we are interested in the production of singly charged scalar of the custodial triplet via $pp\rightarrow tbH_3^{\pm}$.
Next, considering the charged Higgs decay to one gauge boson and one Higgs, this $H_3^{\pm}$ can decay into $W^{\pm}h,\, W^{\pm}H,\, W^{\pm}H_5,\, W^{\mp}H_5^{\pm\pm},\,$ and $ZH_5^{\pm}$.
For the choice of parameter space, it is shown that, the branching ratio of $H_3^{\pm} \rightarrow W^{\pm}H$ is maximum.
Also, following \cite{CMS:2022tkb}, I am interested in singly charged Higgs decay to $W^{\pm}$ and a heavy neutral Higgs further decaying into $\tau^+\tau^-$, and the quintuplet scalars are fermiophobic, my automatic choice is $pp\rightarrow tbH_3^{\pm},\, H_3^{\pm} \rightarrow W^{\pm}H,\, H\rightarrow \tau^+\tau^-$.
To probe this channel, I have also considered the parameter space allowed by the theoretical as well as experimental data, and finally showed that the GM model can accommodate the observed limit obtained by the CMS experiment without constraining the parameter space. 
Besides, I present the possible values of coss-section times branching ratio for the channel $pp\rightarrow tbH_3^{\pm},\, H_3^{\pm} \rightarrow W^{\pm}H,\, H\rightarrow b\overline{b}$ at $\sqrt{s}=14$ TeV.

This letter is arranged as follows.
Section $2$ describes the model briefly, Section $3$ explains the constraints, Section $4$ gives the results, and Section $5$ finally summarizes and concludes.

%
\section{The Georgi-Machacek Model}
\label{model}
\begin{figure}
 \begin{center}
 \includegraphics[width=8.42cm]{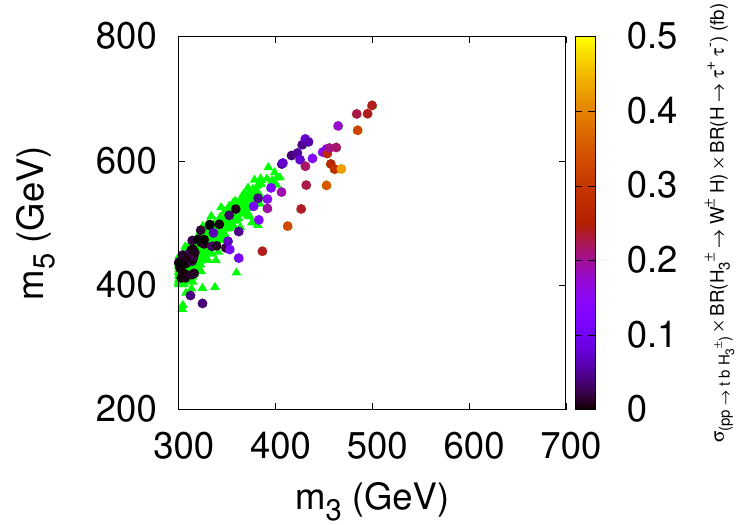} \ \
  \includegraphics[width=8.42cm]{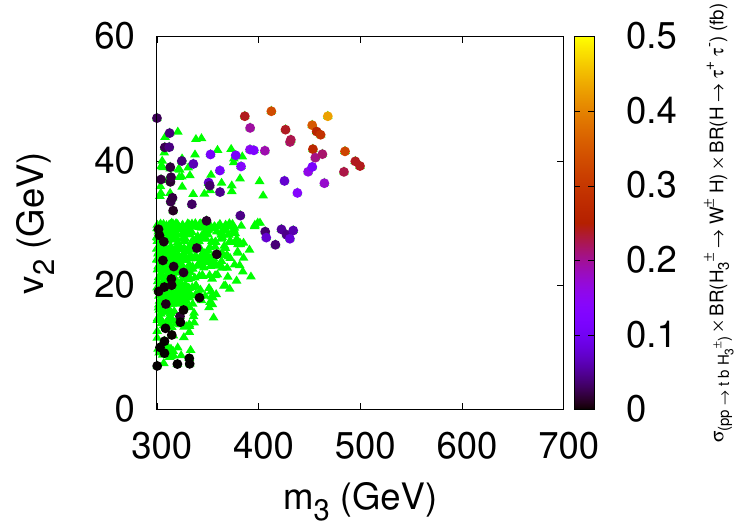} \\
 \includegraphics[width=8.42cm]{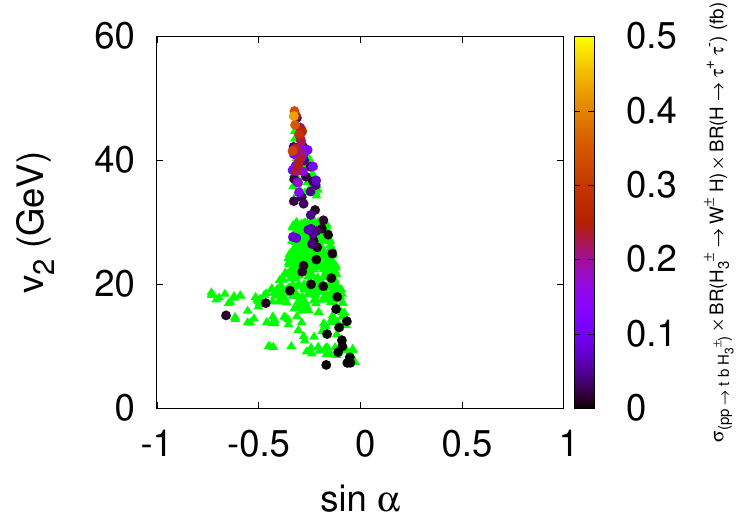} \ \
  \includegraphics[width=8.42cm]{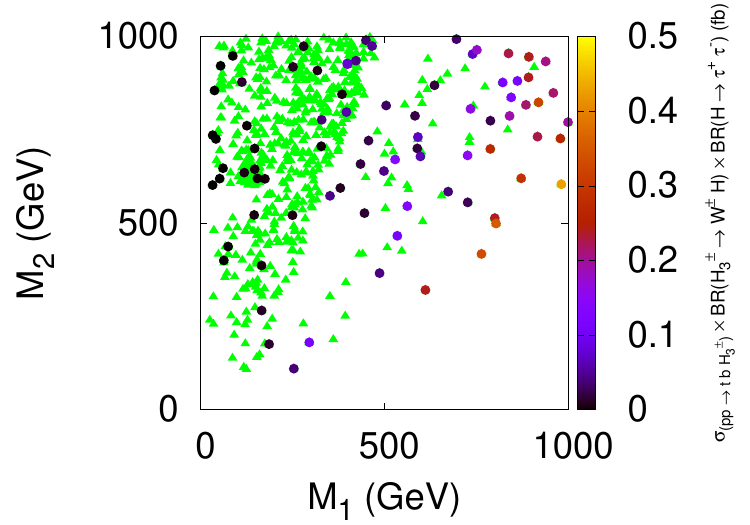} \\
 \end{center}
 \caption{\small Allowed parameter space (green and coloured points) in the $m_3-m_5$,
 $m_3-v_2$, $\sin {\alpha}-v_2$, $M_2-M_1$ plane for $m_h=125$ GeV and $m_H=200$ GeV from theoretical constraints
 and LHC data at $\sqrt{s}=13$ TeV. $\sigma (pp\rightarrow tbH_3^{\pm})\times BR(H_3^{\pm}\rightarrow W^{\pm}H)\times BR(H\rightarrow \tau^+ \tau^-)$ in $fb$ are also shown for points (coloured points except green) for which this value is relatively high. The values of this cross-section times branching ratio are relatively lower for green points, and hence not shown in the plots. }
 \label{fig:overlapped}
 \end{figure}
%
In addition to the SM doublet $\Phi = \left(\phi^+, \phi^0\right)^T$, the scalar sector of the Georgi- Machacek (GM) model consists of two triplets.
Following the notations of \cite{Hartling:2014zca}, the real triplet of hypercharge $Y=0$ and the complex triplet with $Y=2$ are $\xi = \left(\xi^{+},\xi^0,\xi^{-}\right)^T$ and $\chi = \left(\chi^{++},\chi^{+},\chi^0\right)^T$, respectively.

In terms of bi-doublet and bi-triplet,
\be
\Phi =
\begin{pmatrix}
 \phi^{0*} & \phi^{+}\cr
 \phi^{-} & \phi^0
\end{pmatrix}\,
,\quad
X =
\begin{pmatrix}
 \chi^{0*} & \xi^{+} & \chi^{++}\cr
 \chi^{-} & \xi^0 & \chi^{+}\cr
 \chi^{--} & \xi^{-} & \chi^0
\end{pmatrix}\,,
\ee
the scalar potential can be written as,
\bea
V\left(\Phi,X\right) &=& \frac{{\mu_2}^2}{2}\, {\rm Tr}\left(\Phi^\dag\Phi\right)
+ \frac{{\mu_3}^2}{2}\, \rm{Tr}\left(X^\dag X\right)\nonumber\\
&& + {\lambda_1}\left[{\rm Tr}\left(\Phi^\dag\Phi\right)\right]^2
+ {\lambda_2}\, {\rm Tr}\left(\Phi^\dag\Phi\right)\, {\rm Tr}\left(X^\dag X\right) \nonumber\\
&& + {\lambda_3}\, {\rm Tr}\left(X^\dag X X^\dag X\right)
 + {\lambda_4}\left[{\rm Tr}\left(X^\dag X\right)\right]^2\nonumber\\
&&- \frac{\lambda_5}{4}\, {\rm Tr}\left(\Phi^\dag \sigma^a \Phi\sigma^b\right)\, {\rm Tr}\left(X^\dag t^a X t^b\right) \nonumber\\
&& - \frac{M_1}{4} \, {\rm Tr}\left(\Phi^\dag \sigma^a \Phi\sigma^b\right) {\left(U X U^\dag\right)_{ab}}\nonumber\\
&&- {M_2} \, {\rm Tr}\left(X^\dag t^a X t^b\right) {\left(U X U^\dag\right)_{ab}}\, ,
\label{eq:genPot}
\eea
where $\sigma^a$ are the Pauli matrices. The matrices $t^a$s and $U$ are,
\bea
t^1=\frac{1}{\sqrt2}
\begin{pmatrix}
 0 & 1 & 0\cr
 1 & 0 & 1\cr
 0 & 1 & 0
\end{pmatrix}
\,,
t^2=\frac{1}{\sqrt2}
\begin{pmatrix}
 0 & -i & 0\cr
 i & 0 & -i\cr
 0 & i & 0
\end{pmatrix}
\,,\nonumber\\
t^3=
\begin{pmatrix}
 1 & 0 & 0\cr
 0 & 0 & 0\cr
 0 & 0 & -1
\end{pmatrix}\,,
U=
\frac{1}{\sqrt{2}}\begin{pmatrix}
 -1 & 0 & 1\cr
 -i & 0 & -i \cr
 0 & \sqrt{2} & 0
\end{pmatrix}\,.
\eea

To preserve the custodial symmetry, the vevs of the two triplets are the same and equal to $v_2$. The SM doublet vev is equal to $v_1/\sqrt{2}$ leading to $\sqrt{v_1^2+8v_2^2}=v\approx 246$ GeV. I also consider $\tan \beta = 2\sqrt{2}v_2/v_1$.

The scalar sector of GM model possesses ten physical scalars, which can be expressed as,
\bea
&&H_5^{\pm\pm}=\chi^{\pm\pm}\,, \,
H_5^{\pm}=\frac{\left(\chi^{\pm}-\xi^{\pm}\right)}{\sqrt{2}}\,, \,
H_5^0=\sqrt{\frac23}\xi^0-\sqrt{\frac13}\chi^{0R}\,, \nonumber\\
&&H_3^{\pm}=-\sin {\beta}\, \phi^{\pm}+\cos {\beta}\, \frac{\left(\chi^{\pm}+\xi^{\pm}\right)}{\sqrt{2}}\,, \nonumber\\
&&H_3^0=-\sin {\beta}\, \phi^{0I}+\cos {\beta}\, \chi^{0I}\,, \nonumber\\
&&h = \cos {\alpha}\,\, \phi^{0R}-\sin {\alpha}\,\, \sqrt{\frac13}\xi^0+\sqrt{\frac23}\chi^{0R}\,, \nonumber\\
&&H = \sin {\alpha}\,\, \phi^{0R}+\cos {\alpha}\,\, \sqrt{\frac13}\xi^0+\sqrt{\frac23}\chi^{0R}\,.
\label{eq:fields}
\eea
The mass-squared matrices of $h$ and $H$ are given by,
$
{\cal M}^2=
\begin{pmatrix}
 {{\cal M}_{11}}^2 & {{\cal M}_{12}}^2\cr
 {{\cal M}_{21}}^2 & {{\cal M}_{22}}^2
\end{pmatrix}\,,
$
with
${\cal M}_{11}^2 = 8 {\lambda_1}{v_1}^2\,,
{\cal M}_{12}^2 = {\cal M}_{21}^2 = \frac{\sqrt{3}}{2}\left[-M_1+4\left(2\lambda_2-\lambda_5\right)v_2\right]v_1\,,
{\cal M}_{22}^2 = \frac{M_1 v_1^2}{4 v_2}\,.$
One may easily obtain the mixing angle $\alpha$ from
$
\tan{2\alpha}=\frac{2 {{\cal M}_{12}}^2}{{{\cal M}_{22}}^2-{{\cal M}_{11}}^2}.
$\\
$H_5^{\pm\pm,\pm,0}$ have the mass $m_5 = \sqrt{\frac{M_1}{4{v_2}}{v_1}^2+12{M_2}{v_2}+\frac32{\lambda_5}{v_1}^2+8{\lambda_3}{v_2}^2}$, and $H_3^{\pm,0}$ possess the mass $m_3 = \sqrt{\left(\frac{M_1}{4{v_2}}+\frac{\lambda_5}{2}\right)v^2}$. $h$ is considered lighter than $H$, and square of their masses is given by, ${m_{h,H}}^2=\bigg({{\cal M}_{11}}^2+{{\cal M}_{22}}^2\mp\sqrt{\left({{\cal M}_{11}}^2-
{{\cal M}_{22}}^2\right)^2
+4\left({{\cal M}_{12}}^2\right)^2}\bigg)/2$.

%
\section{Theoretical constraints and LHC data}
\label{constraints}
%

 \begin{figure}
  \begin{center}
   \includegraphics[width= 8.42cm]{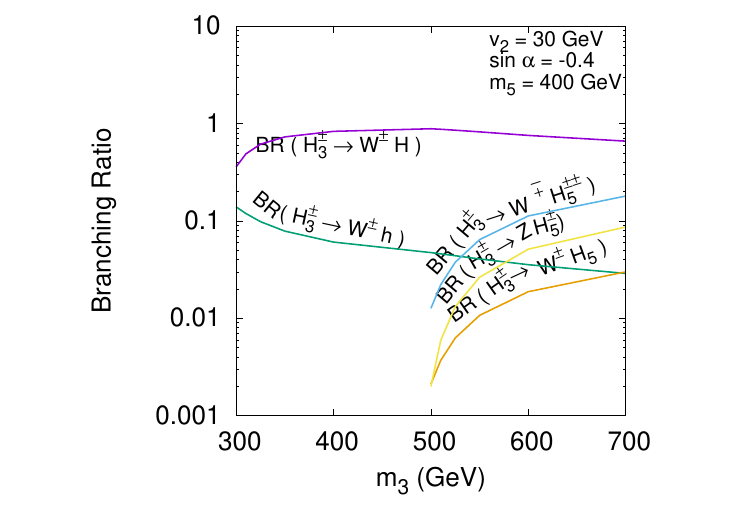} \ \
   \includegraphics[width= 8.42cm]{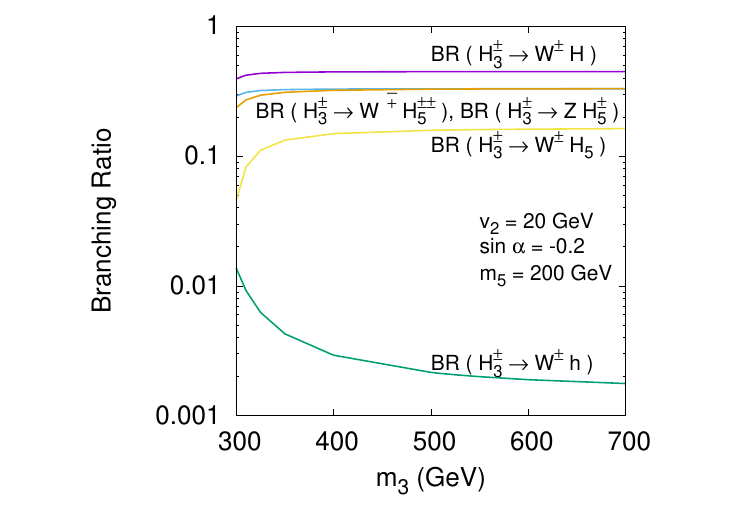} \\
  \end{center}
  \caption{\small Branching ratio of $H_3^{\pm}\rightarrow W^{\pm}h,\,W^{\pm}H,\,W^{\pm}H_5,\,W^{\mp}H_5^{\pm\pm},\,ZH_5^{\pm}$ for $m_H=200$ GeV, $M_1=M_2=$ 100 GeV.}
  \label{fig:Branching}
 \end{figure}

Theoretical constraints in the GM model mainly arise from the perturbative unitarity, electroweak vacuum stability, and the constraints from oblique parameters \cite{Hartling:2014zca, Hartling:2014aga}.
For the constraints coming from the LHC Higgs signal data at $\sqrt{s} = 13$ TeV and the LHC data for the decay of doubly charged Higgs to the same sign $W$ bosons, I follow \cite{CMS:2018lkl,ATLAS:2019slw,CMS:2017fhs}.
Here, I consider the CP-even neutral lighter scalar $h$ as the SM-like Higgs with $m_h\approx125$ GeV.
The coupling modifiers for $h$ are given by,
\be
\kappa_f^h = \frac{v}{v_1}c_{\alpha}\,,\quad
\kappa_V^h = -\frac{1}{3v}(8\sqrt{3}s_{\alpha}v_2\,-\,3 c_{\alpha}v_1).
\ee
For the $h\rightarrow \gamma\gamma$ decay, the charged scalars ($H_5^{\pm\pm}, H_{3,5}^{\pm}$) also take part and hence that is also to be considered in case of the constraints from LHC Higgs data.
Following \cite{CMS:2022tkb}, the mass of the CP-even neutral heavier scalar $H$ is set at $200$ GeV, and the mass $(m_3)$ of the CP-odd scalars varies from $300$ to $700$ GeV.
Also, the mass $(m_5)$ of the CP-even fermiophobic scalars varies from $80$ to $1000$ GeV. 
The other parameters are varied as follows :
$5\le v_2\le 60$ GeV, $-1\le \sin{\alpha}\le 1$, $0\le M_{1,2}\le 1000$ GeV.
The allowed parameter spaces in the $m_3-m_5$, $m_3-v_2$, $\sin {\alpha}-v_2$, $M_2-M_1$ plane are shown in Fig.\ (\ref{fig:overlapped}) as a brief illustration.
In this plot, I also showed the production cross-section times branching ratio of the channel $pp\rightarrow tbH_3^{\pm},\, H_3^{\pm} \rightarrow W^{\pm}H,\, H\rightarrow \tau^+\tau^-$ at $\sqrt{s}=13$ TeV for some allowed parameter points. 
I have not shown the cross-section times branching ratio for the points for which this product is too small.
%
\section{Results}
%

 \begin{figure}
  \begin{center}
 \includegraphics[width=8.42cm]{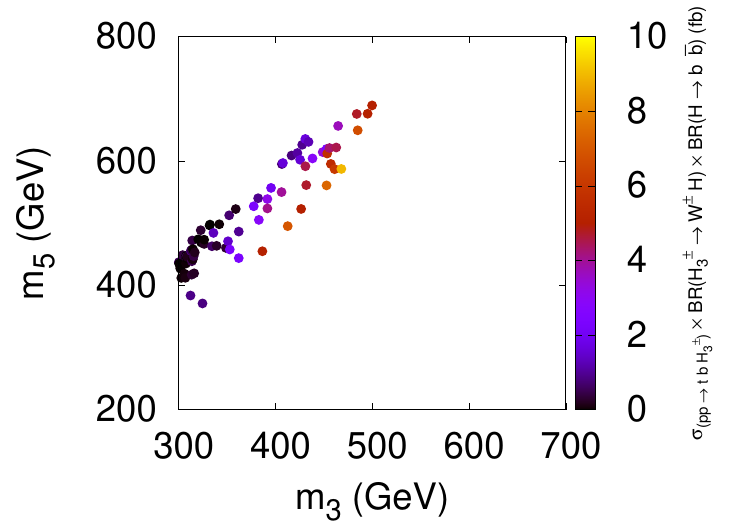} \ \
  \includegraphics[width=8.42cm]{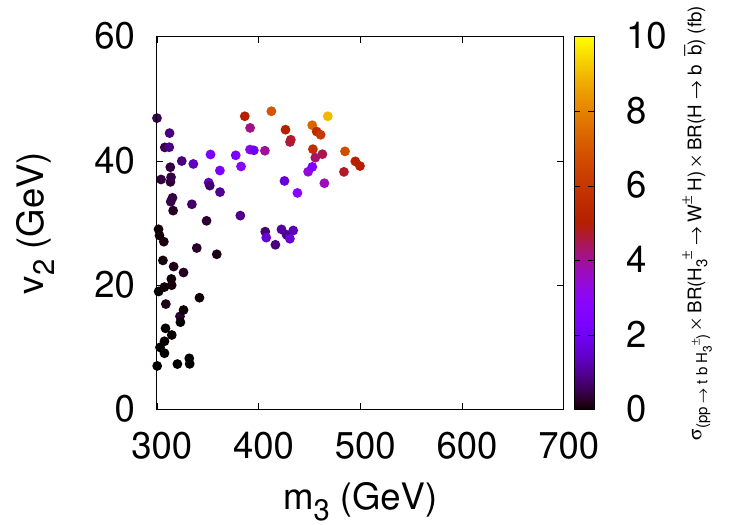} \\
 \includegraphics[width=8.42cm]{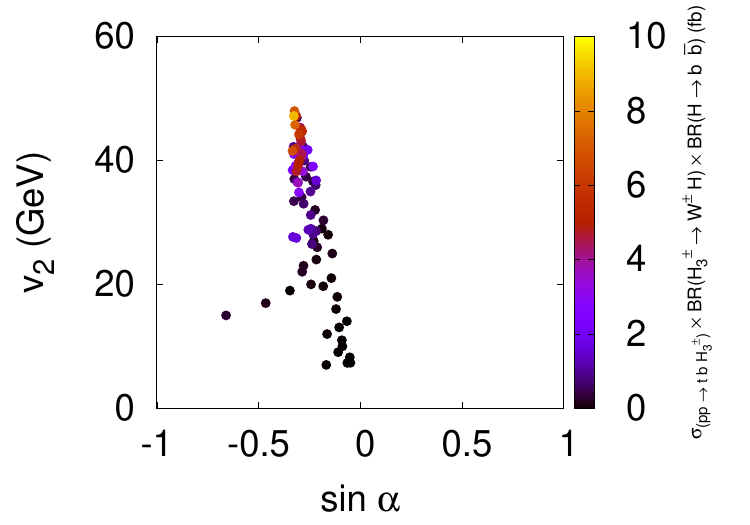} \ \
  \includegraphics[width=8.42cm]{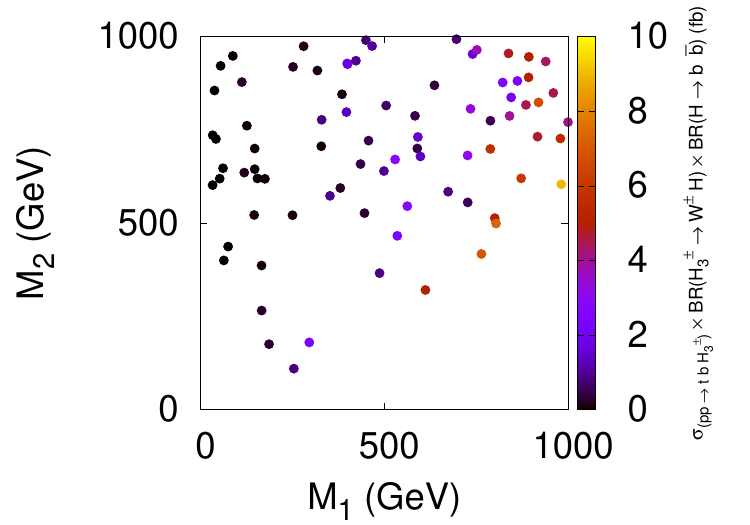} \\
  \end{center}
  \caption{\small Product of cross section and branching ratio for 
  $pp\rightarrow t b H_3^{\pm}, H_3^{\pm}\rightarrow W^{\pm}H, H\rightarrow b\overline{b}$ at $\sqrt{s}=$ 14 TeV for some allowed points in $m_3-m_5,\,m_3-v_2,\,\sin{\alpha}-v_2,\,M_1-M_2$ plane.}
  \label{fig:bb14}
 \end{figure}

This letter focuses on the production and decay of a singly charged Higgs boson decaying into one gauge boson and another Higgs boson followed by the decay of this Higgs boson into two tau leptons following the results obtained in the CMS experiment \cite{CMS:2022tkb}. 
GM model consists of two types of singly charged Higgs $(H_{3,5}^{\pm})$, but the production cross section of $H_3^{\pm}$ is higher than that of $H_5^{\pm}$ through the $pp\rightarrow t b H^{\pm}$ production mechanism. 
The decay of $H_3^{\pm}$ to one gauge boson and one Higgs leads to $H_3^{\pm} \rightarrow W^{\pm} h,\, W^{\pm} H,\, W^{\pm} H_5,\, W^{\mp} H_5^{\pm\pm},\, Z H_5^{\pm}$. 
Fig.\ (\ref{fig:Branching}) shows that for two different choices of the scalar masses, triplet vevs, and mixing angle $\alpha$, the branching ratio of $H_3^{\pm} \rightarrow W^{\pm} H$ is maximum among all other decay channels. 
The sets of parameters in this figure are not consistent with the allowed parameter space, but only for illustration purposes. 
I set $m_H=200$ GeV throughout this work following the reference paper. 

Now, $H$ can decay into a pair of SM as well as BSM particles. 
Following \cite{CMS:2022tkb}, $H\rightarrow \tau^+ \tau^-$ decay is studied here followed by the study of $H\rightarrow b\overline{b}$ as future reference. 
In Fig.\ (\ref{fig:overlapped}), besides the allowed parameter spaces of the GM model (green points), I plot $\sigma (pp\rightarrow tb H_3^{\pm})\times BR(H_3^{\pm} \rightarrow W^{\pm}H) \times BR(H\rightarrow \tau^+ \tau^-)$ in $fb$ at $\sqrt{s}=13$ TeV for the allowed points for which this value is relatively large. 
Next, in Fig.\ (\ref{fig:bb14}), I also showed $\sigma (pp\rightarrow tb H_3^{\pm})\times BR(H_3^{\pm} \rightarrow W^{\pm}H) \times BR(H\rightarrow b \overline{b})$ in $fb$ at $\sqrt{s}=14$ TeV for the allowed points of the GM model. 
The points for which this value is relatively small are not shown in this figure. 
To do these, I implemented the model into FeynRule file \cite{Alloul:2013bka} and obtained the UFO model file for MadGraph \cite{Alwall:2014hca}. 

In the CMS experimental paper, where $\sigma (pp\rightarrow tb H^{\pm})\times BR(H^{\pm} \rightarrow W^{\pm}H) \times BR(H\rightarrow \tau^+ \tau^-)$ in $pb$ is plotted as a function of $m_{H^{\pm}}$ in GeV varying from $300$ to $700$ GeV at $\sqrt{s}=13$ TeV, $m_H=200$ GeV, the production cross-section times the branching ratios have their maximum and minimum value at $0.080$ pb at $m_{H^{\pm}}=300$ GeV and $0.013$ pb at $m_{H^{\pm}}=700$ GeV, respectively. 
From different allowed parameter points, it can be seen that the production cross-section times branching ratios in $pb$ never reach that value, and hence, it can be inferred that the GM model accommodates the experimental results from \cite{CMS:2022tkb} very well. 
Therefore, there is no cut-off in the parameter space from this experimental result.

From different indirect constraints, there are limits on the triplet vev $v_2$ and the strongest one coming from the $b\rightarrow s \gamma$ data \cite{Hartling:2014aga}, which shows that, for the triplet Higgs mass ranging from $300$ to $700$ GeV, the vev of the triplet can go upto $50$ GeV.
Besides, the quintuplet Higgs mass may obtain values as high as $700$ GeV. 
However, these higher values of $v_2$ or $m_5$ do not change our observations and the parameter space of the GM model obtains no constraints from the referred CMS result.

\section{Summary and Conclusions}

The GM model, where the scalar sector of the SM is extended by one real and one complex triplet, preserves the custodial symmetry, and I choose this model to see whether it can accommodate the observed limit of the production cross-section of singly charged Higgs boson with mass $300-700$ GeV times its decay to $W^{\pm}$ and a heavy neutral Higgs with mass 200 GeV, which further decays into a pair of tau leptons as studied in \cite{CMS:2022tkb}.
The scalar sector of the GM model contains ten physical
scalars, where four are neutral $(h,\,H,\,H_3,\,H_5)$, four are singly charged $(H_{3,5}^{\pm})$, and two are doubly charged $(H_5^{\pm\pm})$.
Among the singly charged particles, the cross-section of $H_3^{\pm}$ is higher, and the decay branching ratio of this to $W^{\pm}H$ is maximum among other possible decays of $H_3^{\pm}$ to one gauge boson and one Higgs boson. Therefore, this letter studies the process $pp\rightarrow tbH_3^{\pm} \rightarrow tbW^{\pm}H \rightarrow tbW^{\pm}\tau^+\tau^-$ for different parameter choices of the GM model and shows that this model can accommodate the observed limit obtained from the CMS experimental result.
However, any future modification from the LHC data for this channel can change this inference and also can probe the parameter space of the GM model. 
This letter also studies the process $pp\rightarrow tbH_3^{\pm} \rightarrow tbW^{\pm}H \rightarrow tbW^{\pm}b\overline{b}$ for the allowed parameter points and predicts the corresponding cross-section times branching ratio for future reference.

\vspace{0.5cm}
{\em{\bf Acknowledgements}} --- The author acknowledges the Department of Science and Technology, Government of India for financial support through the SERB-NPDF scholarship with grant no. PDF/2022/001784, and Professor Anirban Kundu and Professor Tirtha Sankar Ray for valuable discussions.


 
\end{document}